\documentclass[aps,prl,superscriptaddress,showpacs,floatfix,twocolumn]{revtex4}

\usepackage{graphicx}	

\begin{document}

\newcommand{\pt}{$p_T$ }
\newcommand{\gevc}{GeV/$c$ }
\newcommand{\flow}{$v_2$ }

\title{Correlated Production of $p$ and $\bar{p}$ in Au+Au Collisions at 
$\sqrt{s_{NN}}$~=~200~GeV}

\newcommand{\abilene}{Abilene Christian University, Abilene, TX 79699, U.S.}
\newcommand{\banaras}{Department of Physics, Banaras Hindu University, Varanasi 221005, India}
\newcommand{\bnl}{Brookhaven National Laboratory, Upton, NY 11973-5000, U.S.}
\newcommand{\caucr}{University of California - Riverside, Riverside, CA 92521, U.S.}
\newcommand{\charlesczech}{Charles University, Ovocn\'{y} trh 5, Praha 1, 116 36, Prague, Czech Republic}
\newcommand{\ciae}{China Institute of Atomic Energy (CIAE), Beijing, People's Republic of China}
\newcommand{\cns}{Center for Nuclear Study, Graduate School of Science, University of Tokyo, 7-3-1 Hongo, Bunkyo, Tokyo 113-0033, Japan}
\newcommand{\colorado}{University of Colorado, Boulder, CO 80309, U.S.}
\newcommand{\columbia}{Columbia University, New York, NY 10027 and Nevis Laboratories, Irvington, NY 10533, U.S.}
\newcommand{\czechtech}{Czech Technical University, Zikova 4, 166 36 Prague 6, Czech Republic}
\newcommand{\dapnia}{Dapnia, CEA Saclay, F-91191, Gif-sur-Yvette, France}
\newcommand{\debrecen}{Debrecen University, H-4010 Debrecen, Egyetem t{\'e}r 1, Hungary}
\newcommand{\elte}{ELTE, E{\"o}tv{\"o}s Lor{\'a}nd University, H - 1117 Budapest, P{\'a}zm{\'a}ny P. s. 1/A, Hungary}
\newcommand{\fit}{Florida Institute of Technology, Melbourne, FL 32901, U.S.}
\newcommand{\fsu}{Florida State University, Tallahassee, FL 32306, U.S.}
\newcommand{\gsu}{Georgia State University, Atlanta, GA 30303, U.S.}
\newcommand{\hiroshima}{Hiroshima University, Kagamiyama, Higashi-Hiroshima 739-8526, Japan}
\newcommand{\ihepprot}{IHEP Protvino, State Research Center of Russian Federation, Institute for High Energy Physics, Protvino, 142281, Russia}
\newcommand{\illuiuc}{University of Illinois at Urbana-Champaign, Urbana, IL 61801, U.S.}
\newcommand{\instpasczech}{Institute of Physics, Academy of Sciences of the Czech Republic, Na Slovance 2, 182 21 Prague 8, Czech Republic}
\newcommand{\isu}{Iowa State University, Ames, IA 50011, U.S.}
\newcommand{\jinrdubna}{Joint Institute for Nuclear Research, 141980 Dubna, Moscow Region, Russia}
\newcommand{\kaeri}{KAERI, Cyclotron Application Laboratory, Seoul, South Korea}
\newcommand{\kek}{KEK, High Energy Accelerator Research Organization, Tsukuba, Ibaraki 305-0801, Japan}
\newcommand{\kfki}{KFKI Research Institute for Particle and Nuclear Physics of the Hungarian Academy of Sciences (MTA KFKI RMKI), H-1525 Budapest 114, POBox 49, Budapest, Hungary}
\newcommand{\korea}{Korea University, Seoul, 136-701, Korea}
\newcommand{\kurchatov}{Russian Research Center ``Kurchatov Institute", Moscow, Russia}
\newcommand{\kyoto}{Kyoto University, Kyoto 606-8502, Japan}
\newcommand{\labllr}{Laboratoire Leprince-Ringuet, Ecole Polytechnique, CNRS-IN2P3, Route de Saclay, F-91128, Palaiseau, France}
\newcommand{\lawllnl}{Lawrence Livermore National Laboratory, Livermore, CA 94550, U.S.}
\newcommand{\losalamos}{Los Alamos National Laboratory, Los Alamos, NM 87545, U.S.}
\newcommand{\lpc}{LPC, Universit{\'e} Blaise Pascal, CNRS-IN2P3, Clermont-Fd, 63177 Aubiere Cedex, France}
\newcommand{\lund}{Department of Physics, Lund University, Box 118, SE-221 00 Lund, Sweden}
\newcommand{\muenster}{Institut f\"ur Kernphysik, University of Muenster, D-48149 Muenster, Germany}
\newcommand{\myongji}{Myongji University, Yongin, Kyonggido 449-728, Korea}
\newcommand{\nagasaki}{Nagasaki Institute of Applied Science, Nagasaki-shi, Nagasaki 851-0193, Japan}
\newcommand{\newmex}{University of New Mexico, Albuquerque, NM 87131, U.S. }
\newcommand{\nmsu}{New Mexico State University, Las Cruces, NM 88003, U.S.}
\newcommand{\ornl}{Oak Ridge National Laboratory, Oak Ridge, TN 37831, U.S.}
\newcommand{\orsay}{IPN-Orsay, Universite Paris Sud, CNRS-IN2P3, BP1, F-91406, Orsay, France}
\newcommand{\peking}{Peking University, Beijing, People's Republic of China}
\newcommand{\pnpi}{PNPI, Petersburg Nuclear Physics Institute, Gatchina, Leningrad region, 188300, Russia}
\newcommand{\riken}{RIKEN, The Institute of Physical and Chemical Research, Wako, Saitama 351-0198, Japan}
\newcommand{\rikjrbrc}{RIKEN BNL Research Center, Brookhaven National Laboratory, Upton, NY 11973-5000, U.S.}
\newcommand{\rikkyo}{Physics Department, Rikkyo University, 3-34-1 Nishi-Ikebukuro, Toshima, Tokyo 171-8501, Japan}
\newcommand{\saispbstu}{Saint Petersburg State Polytechnic University, St. Petersburg, Russia}
\newcommand{\saopaulo}{Universidade de S{\~a}o Paulo, Instituto de F\'{\i}sica, Caixa Postal 66318, S{\~a}o Paulo CEP05315-970, Brazil}
\newcommand{\seoulnat}{System Electronics Laboratory, Seoul National University, Seoul, South Korea}
\newcommand{\stonybrkc}{Chemistry Department, Stony Brook University, Stony Brook, SUNY, NY 11794-3400, U.S.}
\newcommand{\stonycrkp}{Department of Physics and Astronomy, Stony Brook University, SUNY, Stony Brook, NY 11794, U.S.}
\newcommand{\subatech}{SUBATECH (Ecole des Mines de Nantes, CNRS-IN2P3, Universit{\'e} de Nantes) BP 20722 - 44307, Nantes, France}
\newcommand{\tenn}{University of Tennessee, Knoxville, TN 37996, U.S.}
\newcommand{\titech}{Department of Physics, Tokyo Institute of Technology, Oh-okayama, Meguro, Tokyo 152-8551, Japan}
\newcommand{\tsukuba}{Institute of Physics, University of Tsukuba, Tsukuba, Ibaraki 305, Japan}
\newcommand{\vandy}{Vanderbilt University, Nashville, TN 37235, U.S.}
\newcommand{\waseda}{Waseda University, Advanced Research Institute for Science and Engineering, 17 Kikui-cho, Shinjuku-ku, Tokyo 162-0044, Japan}
\newcommand{\weizmann}{Weizmann Institute, Rehovot 76100, Israel}
\newcommand{\yonsei}{Yonsei University, IPAP, Seoul 120-749, Korea}
\affiliation{\abilene}
\affiliation{\banaras}
\affiliation{\bnl}
\affiliation{\caucr}
\affiliation{\charlesczech}
\affiliation{\ciae}
\affiliation{\cns}
\affiliation{\colorado}
\affiliation{\columbia}
\affiliation{\czechtech}
\affiliation{\dapnia}
\affiliation{\debrecen}
\affiliation{\elte}
\affiliation{\fit}
\affiliation{\fsu}
\affiliation{\gsu}
\affiliation{\hiroshima}
\affiliation{\ihepprot}
\affiliation{\illuiuc}
\affiliation{\instpasczech}
\affiliation{\isu}
\affiliation{\jinrdubna}
\affiliation{\kaeri}
\affiliation{\kek}
\affiliation{\kfki}
\affiliation{\korea}
\affiliation{\kurchatov}
\affiliation{\kyoto}
\affiliation{\labllr}
\affiliation{\lawllnl}
\affiliation{\losalamos}
\affiliation{\lpc}
\affiliation{\lund}
\affiliation{\muenster}
\affiliation{\myongji}
\affiliation{\nagasaki}
\affiliation{\newmex}
\affiliation{\nmsu}
\affiliation{\ornl}
\affiliation{\orsay}
\affiliation{\peking}
\affiliation{\pnpi}
\affiliation{\riken}
\affiliation{\rikjrbrc}
\affiliation{\rikkyo}
\affiliation{\saispbstu}
\affiliation{\saopaulo}
\affiliation{\seoulnat}
\affiliation{\stonybrkc}
\affiliation{\stonycrkp}
\affiliation{\subatech}
\affiliation{\tenn}
\affiliation{\titech}
\affiliation{\tsukuba}
\affiliation{\vandy}
\affiliation{\waseda}
\affiliation{\weizmann}
\affiliation{\yonsei}
\author{A.~Adare}	\affiliation{\colorado}
\author{S.~Afanasiev}	\affiliation{\jinrdubna}
\author{C.~Aidala}	\affiliation{\columbia}
\author{N.N.~Ajitanand}	\affiliation{\stonybrkc}
\author{Y.~Akiba}	\affiliation{\riken} \affiliation{\rikjrbrc}
\author{H.~Al-Bataineh}	\affiliation{\nmsu}
\author{J.~Alexander}	\affiliation{\stonybrkc}
\author{A.~Al-Jamel}	\affiliation{\nmsu}
\author{K.~Aoki}	\affiliation{\kyoto} \affiliation{\riken}
\author{L.~Aphecetche}	\affiliation{\subatech}
% ...
\collaboration{PHENIX Collaboration} \noaffiliation

\begin{abstract}
Correlations between $p$ and $\bar{p}$ at transverse momenta typical of
enhanced baryon production in Au+Au collisions are reported.
The PHENIX experiment measures same and opposite sign baryon pairs
in Au+Au collisions at $\sqrt{s_{NN}}$~=~200~GeV.
Correlated production of $p$ and $\bar{p}$  with the trigger 
particle from the range 2.5~$<p_T<$~4.0~\gevc and
the associated particle with 1.8~$<p_T<$~2.5~\gevc is observed 
to be nearly independent of the centrality of the collisions. 
Same sign pairs show no correlation at any centrality.  
The conditional yield of mesons triggered by baryons (and anti-baryons) 
and mesons in the same \pt 
range rises with increasing centrality, except for 
the most central collisions, where baryons
show a significantly smaller number of  associated mesons. 
These data are consistent with a picture in which hard scattered partons
produce correlated $p$ and $\bar{p}$ in  the \pt region of the baryon
excess.
\end{abstract}

\date{\today}

	
\maketitle

\section{Introduction}
A remarkable feature of relativistic heavy ion collisions
at RHIC energies is the enhanced production of baryons and anti-baryons
relative to mesons  at intermediate transverse momenta 
(2~$<p_T<$~5~GeV/$c$)~\cite{ppg015,ppg026}. 
In central Au+Au collisions at $\sqrt{s_{NN}}$~=200~GeV the baryon/meson 
ratio is a factor of three higher than in $p+p$ collisions, while  in peripheral 
Au+Au collisions and in d+Au collisions~\cite{ppg030} at the same 
energy only a small increase ($<$~20\%) is observed. 
The production of protons and anti-protons at intermediate \pt 
in Au+Au collisions  scales with the number of 
binary nucleon-nucleon collisions~\cite{ppg015}, contrary to the 
suppression of pion production~\cite{ppg014}. Similar behavior has 
been observed for strange baryons ($\Lambda$ and $\bar{\Lambda}$) 
and mesons ($K_S^0$)~\cite{starlambda}.

In this same momentum range in $p+p$ collisions at the same center of mass energy the 
dominant production mechanism shifts from soft processes characterized by 
non-perturbative low momentum transfer
scattering to hard scattering processes characterized
by large momentum transfer parton-parton scattering followed
by fragmentation of the scattered partons into final state
hadrons. Pion production in  $\sqrt{s_{NN}}$=200~GeV $p+p$ collisions is reasonably 
well described by perturbative QCD (pQCD) down to $p_T\approx$~2~GeV \cite{ppg024}.
There are large variations in the $p$ and $\bar{p}$ yield
 among various fragmentation
functions which make it difficult to establish a definite pQCD 
expectation for the $p$ and $\bar{p}$ spectra in p+p collisions~\cite{starxt}.
Another estimate of the transition from hard to soft physics
can be obtained from the $x_T=2p_T/\sqrt{s}$ scaling of the single particle 
cross sections.
The cross section can be written as~\cite{berman,blankenbecler}:
\begin{equation}
E\frac{d^3\sigma}{dp^3}=\frac{1}{\sqrt{s}^{n(\sqrt{s},x_T)}}G(x_T)
\label{eqxt}
\end{equation}
At high $x_T$ the value of $n$ is found to be independent of
both $\sqrt{s}$ and $x_T$.  Since the power of $n$ is related
to the quantum exchanged and the number of point-like scatterers, 
the $x_T$ region corresponding to
the asymptotic $n$ value is understood be to the region where
particle production is dominated by hard scattering.
Recent measurements of the 
$p$ and $\bar{p}$ and $\pi^{\pm}$ spectra show the cross section can be
described by consistent values of $n$ for $p_T>$~2~\gevc
for both the $\pi^\pm$ ($n$~=~6.8~$\pm$~0.5) and $p$ and $\bar{p}$ 
($n$~=~6.5~$\pm$~1.0)~\cite{starxt}
indicating, together with the
agreement between the data and pQCD calculations at high $p_T$~\cite{ppg024,starxt},
 that at $\sqrt{s}$=200~GeV the transition from hard
to soft particle production happens at $p_T\approx$~2~\gevc.
Since the fragmentation process is believed to be independent of 
center of mass energy or collision system, 
baryon and anti-baryon production in central Au+Au collisions 
appears inconsistent with hard-scattering followed by a universal fragmentation. 

The theoretical models that are successful in reproducing the measured 
single particle spectra, 
baryon/meson ratios and the nuclear modification factors usually 
invoke some mechanism to extend the 
range of soft particle production for baryons to higher \pt than
that for mesons. This is either done 
based on the particle mass (in hydrodynamics 
models~\cite{Teaney,Hirano}) or on the quark content 
(quark recombination models~\cite{grecoprc,friesprc,hwa1}).
An alternative approach, which involves 
production of baryons through gluon junctions has also been shown to 
reproduce the data~\cite{topor-pop}. 
In this paper we study two-particle angular correlations involving 
$p$, $\bar{p}$, and mesons ($\pi^{\pm}$, $K^{\pm}$).
 This approach gives information about the hadron production in hard-scattering 
processes, which is inaccessible 
from single particle measurements.        

Previous studies~\cite{starb2b,ppg033,ppg032} in Au+Au collisions 
show that at intermediate \pt  
particles are correlated in azimuthal angle in a manner consistent with jet
fragmentation. Namely, particles are emitted close together 
when they come from  
fragmentation of the same jet (near side correlations) 
or approximately 
back-to-back when they come  from the fragmentation of the associated 
di-jet (away side correlations).
Strong modifications of the yields, shapes  and particle composition of these
correlations are seen from peripheral to central Au+Au collisions.
The yields are quantified by the number 
of associated particles per trigger (conditional yield)  after the combinatorial
background from the underlying event has been subtracted.
The conditional yield is measured separately for the near
and away side correlations.
The near side conditional yield  
increases~\cite{starb2b,ppg033,ppg032} with centrality and the away side
shape has a peak at $\Delta\phi\approx$~2~rad and no
peak at $\Delta\phi\approx\pi$~rad.
Correlations between identified baryons ($p$, $\bar{p}$)
and charged hadrons~\cite{ppg033}  rule out baryon production 
from 2.5~$<p_T<$~4.0~\gevc  as coming dominantly 
from a thermal parton source with no correlations.
Moreover, the magnitude of baryon and meson
triggered correlations with other
charged hadrons in the same event is similar,
indicating that the baryon excess in the intermediate \pt range is
associated with hard parton-parton scattering.

There are two main recombination models which have attempted to 
address the connection between hard scattering and recombination.  The
model of Hwa and Yang \cite{hwa1} has calculated that fragmenting
partons from hard scattering process have a high probability to 
recombine with thermal quarks from the medium.  The model of Fries 
{\it et al.}~\cite{friescs}
finds these effects to be unimportant, but has correlations from
fast partons losing energy in the medium creating a region around
the parton trajectory
with a slightly increased temperature and with additional momentum
in the direction of the energetic parton.
Partons from this region then recombine into hadrons which are
correlated with the fast parton direction and with each other.

In order to further explore the jet-like structure of the baryon
excess, here we present results on the  angular correlations between
two identified particles.   The baryon production mechanism is
studied via correlations between two charge separated $p$ and $\bar{p}$.
If main the source of the baryon excess is jets that fragment outside
the medium, the charge dependence of correlations between
$p$ and $\bar{p}$ should be the same from peripheral collisions,
where baryon production at intermediate \pt is nearly
unmodified from p+p collisions, to central collisions.
A centrality dependence of the  charge combinations of $p$ and $\bar{p}$
correlations would provide
evidence for novel baryon production scenarios in central Au+Au collisions.

Correlations between baryon and meson triggers with associated mesons 
are studied as well.  
Meson trigger-associated mesons (meson-meson) correlations
provide a baseline for jet fragmentation in Au+Au collisions.
Separating the baryon-hadron correlations from \cite{ppg033} into
baryon-meson and baryon-baryon correlations allows greater sensitivity
to possible recombination effects.  The recombination model of Fries {\it et al.} 
\cite{friescs} predicts a greater amplification of baryon-baryon
correlations relative to baryon-meson correlations on
the near side because of the larger number
of possible correlations between the valence quarks.  

Near side
correlations from jets in p+p 
collisions are observed to be balanced by away side
correlations from the associated di-jet \cite{ppg029}.  The away side
correlations at intermediate \pt in Au+Au collisions have been shown to have 
a modified shape in Au+Au collisions~\cite{ppg032}.
Here we measure the away side correlations with both particles identified
in order to see if the  dependence on the trigger and associated particle type
changes with centrality. 
If the baryon and meson triggers are from hard scattering with approximately the
same momentum transfer, we expect the away side correlations to be independent
of the trigger particle type. 
%

The paper is organized as follows: Section ~\ref{section:experimental} describes the 
experimental method and setup, the results are presented in Section~\ref{section:results}, and 
Section~\ref{section:discussion} is devoted to discussion.    

\section{Experimental procedure and setup}
\label{section:experimental}
Two particle correlations have been widely used to
study jets in heavy ion collisions 
\cite{starb2b,ppg033,ppg032,starlowpt,stardijet} where, due to the high
multiplicity and moderate jet energy the direct reconstruction 
of jets by standard algorithms, utilizing hadronic calorimetry and 
cluster algorithms, is not yet possible.
In this approach particles are divided into two classes,
{\it triggers} and {\it associated particles}.
We classify the triggers and associated particles by their $p_T$, particle type and charge.
The triggers have 2.5~$<p_T<$ ~4.0~\gevc and the associated particles have 1.8~$<p_T<$~2.5~\gevc. 
Thus both particles originate from a region in \pt that is  
consistent with hard scattering in $p+p$ collisions and shows an excess of baryons
relative to mesons in Au+Au collisions. A distribution of the azimuthal angular difference 
$\Delta\phi$ between trigger-associated particle pairs is constructed and normalized by
 the number of triggers.

The data presented here are based on an analysis of 600M Au+Au
events collected by the PHENIX experiment in 2004 with a minimum bias trigger.  
Charged particles are reconstructed in the central arms of PHENIX 
using a combination of drift chambers and one layer of multi-wire proportional 
chamber with pad readout (PC1)~\cite{trackingnim}, each covering $\Delta \phi = \pi/2$ in azimuthal angle
 and $|\eta| <0.35$ in pseudorapity. The pattern recognition is based on a combinatorial Hough transform
in the track bend plane.  The polar  angle is determined  from the hit position in PC1 and the 
collision vertex along the beam axis measured by the Beam-Beam Counters (BBC).
The BBC are positioned at $|\eta|=$3-4.
Particles are identified by their mass calculated from the measured
 momentum and time-of-flight information. 
The global start time is provided by BBC, while the stop time is measured 
by the PHENIX high resolution 
time of flight detector (TOF) or the lead-scintillator electromagnetic 
calorimeter (EMCal) which 
provide a 4$\sigma$ $K/p$ separation up to \pt $\approx$~4.0~GeV/$c$ and  
\pt $\approx$~2.5~GeV/$c$, respectively. The trigger particles 
are identified in the TOF, which covers a 
portion of the PHENIX East arm ($\Delta \phi = \pi/4$). The associated particles are 
identified in either the EMCal or the TOF, which together cover the 
entire PHENIX azimuthal acceptance.
For both triggers and associated particles a 2$\sigma$ spatial match is required between
the track projection and the hit position in the particle identification
detector.  Monte Carlo studies have shown that,
due to the decay kinematics in the trigger and partner \pt range used here, 
the contribution from 
$\Lambda\to p\pi^-$ and $\bar{\Lambda}\to \bar{p}\pi^+$ resonance decays,
which could produce correlations mimicking the jet signal, is
negligible.

We perform a correction for the non-uniform pair acceptance
in $\Delta\phi$ in PHENIX. This correction is constructed by measuring the 
$\Delta\phi$ distribution from trigger-associated particle pairs where each
particle comes from a different event.  Dividing the same-event by the 
mixed-event distribution removes the effects of the PHENIX acceptance and leaves
only the true correlations.  The multiplicity of the combinatorial
background of the underlying event is determined absolutely by the 
convolution of the measured trigger and associated particle single particle rates with
an additional correction for centrality 
correlations~\cite{ppg033} which raises the combinatorial
background level by $\approx$~0.2\% in the most central collisions
and $\approx$~25\% in peripheral collisions.
A correction for the associated particle reconstruction
efficiency and acceptance is applied by matching
the observed rates for the corresponding single 
particle spectra  measured in~\cite{ppg026}.
No correction has been made for
$p$ and $\bar{p}$ originating from weak decays of $\Lambda$ and 
$\bar{\Lambda}$; approximately 30\% of the 
measured $p$ and $\bar{p}$ are from these decays~\cite{ppg026}.
The PHENIX $\eta$ acceptance is narrow compared
to the width of the typical jet cone in p+p collisions \cite{ppg029}, 
so we do not measure the entire conditional yield associated
with the trigger particle in the associated particle \pt range.
The results are reported for both trigger and associated particles
 within $|\eta|<$~0.35 without 
extrapolating in pseudorapidity. The centrality dependence of the conditional 
yields allows us to quantify changes in the jet-like correlations as a function
of centrality and particle type, despite the limited acceptance.

\begin{table*}
\caption{The \flow values and statistical and systematic errors for the centrality and
\pt bins used in the analysis.}
\label{v2tab}
\begin{ruledtabular} \begin{tabular}{cccc}
\hline
 & \multicolumn{3}{c}{\flow Values $\pm$ Statistical Error $\pm$ Systematic Error}\\
 & Triggers & \multicolumn{2}{c}{Associated Particles}\\
Centrality & 2.5$<p_T<$4.0~\gevc & 1.8$<p_T<$2.0~\gevc & 2.0$<p_T<$2.5~\gevc\\ \hline
Baryons\\
0-5\% & 0.083 $\pm$ 0.006 $\pm$ 0.017 &  0.064 $\pm$ 0.004 $\pm$ 0.013&  0.068 $\pm$ 0.004 $\pm$ 0.014\\
5-10\% & 0.126 $\pm$ 0.005 $\pm$ 0.016 &  0.089 $\pm$ 0.003 $\pm$ 0.011&   0.108 $\pm$ 0.003 $\pm$ 0.013\\
10-20\% & 0.176 $\pm$ 0.003 $\pm$ 0.013&  0.134 $\pm$ 0.002 $\pm$ 0.010&   0.154 $\pm$ 0.002 $\pm$ 0.011\\
20-40\% & 0.234 $\pm$ 0.003 $\pm$ 0.014&  0.182 $\pm$ 0.002 $\pm$ 0.011&  0.203 $\pm$ 0.002 $\pm$ 0.012\\
40-60\%& 0.264 $\pm$ 0.006 $\pm$ 0.015&  0.211 $\pm$ 0.004 $\pm$ 0.012&   0.235 $\pm$ 0.004 $\pm$ 0.013\\
60-90\% & 0.276 $\pm$ 0.033 $\pm$ 0.044&  0.158 $\pm$ 0.023 $\pm$ 0.025&  0.179 $\pm$ 0.021 $\pm$ 0.029\\
Mesons\\
0-5\% & 0.072 $\pm$ 0.007 $\pm$ 0.015&  0.067 $\pm$ 0.003 $\pm$ 0.014&  0.078 $\pm$ 0.003 $\pm$ 0.016\\
5-10\% & 0.109 $\pm$ 0.005 $\pm$ 0.014 &  0.102 $\pm$ 0.002 $\pm$ 0.013&  0.103 $\pm$ 0.002 $\pm$ 0.013\\
10-20\% & 0.142 $\pm$ 0.003 $\pm$ 0.011&  0.133 $\pm$ 0.001 $\pm$ 0.010&  0.140 $\pm$ 0.001 $\pm$ 0.010\\
20-40\% & 0.185 $\pm$ 0.003 $\pm$ 0.011&  0.172 $\pm$ 0.001 $\pm$ 0.010&  0.180 $\pm$ 0.001 $\pm$ 0.011\\
40-60\% & 0.186 $\pm$ 0.006 $\pm$ 0.010&  0.188 $\pm$ 0.003 $\pm$ 0.011&  0.191 $\pm$ 0.003 $\pm$ 0.011\\
60-90\% & 0.178 $\pm$ 0.027 $\pm$ 0.029&  0.173 $\pm$ 0.013 $\pm$ 0.028&  0.172 $\pm$ 0.014 $\pm$ 0.28\\
\\
\end{tabular} \end{ruledtabular}
\end{table*}

Elliptic flow is an azimuthal correlation between particles
due to the anisotropy in the initial collision geometry.
This  angular correlation is unrelated to jet fragmentation and thus 
produces a background for this measurement.
The correlations due to elliptic flow are removed by modulating the azimuthally uniform combinatorial background
by $1+2v_2^{trig}v_2^{part}\cos(\Delta\phi)$ where
$v_2^{trig}$ and $v_2^{part}$ represent the strength of the elliptic flow signal for
the trigger and associated particle, respectively. 
The \flow parameter is defined by the 2nd harmonic azimuthal anisotropy, 
\flow = $\left<\cos[2({\phi-\Psi)]}\right>$, where $\phi$ is the azimuthal angle of 
emitted particle, $\Psi$ is the azimuthal angle of event plane in a given collision, 
and the bracket denotes the average over all particles and events~\cite{eventplane}.
We measure \flow of charged baryons and mesons
 at mid-rapidity, $|\eta| < 0.35$, for each centrality and 
\pt bin through the event plane method~\cite{eventplane}. The azimuthal angle of the event plane is 
determined by the BBC  using the elliptic moment definition \cite{ppg022}. 
The large rapidity difference,
 $|\Delta\eta| \sim 3$, between the central arms and the BBCs helps to reduce the   
non-flow contributions to the measured \flow, especially those arising from 
di-jets. The systematic errors on the \flow value are dominated by the uncertainty 
in the correction for the event plane resolution~\cite{ppg022}.
The \flow values used in this analysis are shown in Table \ref{v2tab}. 
They are consistent with prior 
PHENIX \flow measurements~\cite{ppg022} in the common centrality bins.
The analysis has been performed separately for
associated particles in two transverse momentum ranges: 1.8~$<p_T<$~2.0~\gevc and 2.0~$<p_T<$~2.5~\gevc.
This minimizes effects due to the variation of \flow over
the width of the associated particle \pt bin.
In order to minimize the statistical errors, the results shown here are a sum of the two bins.

The systematic errors on the conditional yields
are due to the systematic and statistical uncertainties
on the \flow values, the uncertainty in the corrections 
for the centrality correlations in the combinatorial background
and in the centrality dependence of the efficiency corrections.
The systematic error on the centrality dependence of the
efficiency corrections is 6\% for meson associated particles and 5\% for
inclusive $p$, $\bar{p}$ and baryon ($p$ and $\bar{p}$ combined) 
associated particles independent of centrality.
The size of the systematic error on the conditional yield attributed to a systematic 
uncertainty in the elliptic flow determination is largest in the most central collisions.
The systematic error on the centrality correlations is $\approx$~60\%
of the correction in central collisions and $\approx$~5\% of the correction
in peripheral collisions.
There are additional systematic errors which are not shown in the figures
 in Section ~\ref{section:results} that come from the centrality independent 
normalization of the efficiency corrections and move all points with the same associated particle type together.
These are 8.9\% for $p$ and $\bar{p}$ associated particles,  11.4\% for baryon ($p$
and $\bar{p}$) associated particles, and 13.6\% for meson associated particles.

\section{Results}
\label{section:results}

Our goal is to study the jet contribution to baryon and anti-baryon production at intermediate \pt 
where an excess of baryons over mesons is observed. Thus we choose 
trigger baryons from the 
range 2.5~ $<p_T <$~ 4.0~GeV/$c$ and associated particles in the 
range 1.8 ~$<p_T <$~2.5~GeV/$c$ and construct 
two-particle azimuthal correlation distributions. 
With the larger  data sample obtained in 2004 we 
are able to extend our previous studies~\cite{ppg033} by studying
 the proton and anti-proton triggers, as 
well as identifying the associated particles. Several combinations of trigger-associated 
particle types are presented below.

\begin{figure*}[thb]
\includegraphics[width=0.7\linewidth]{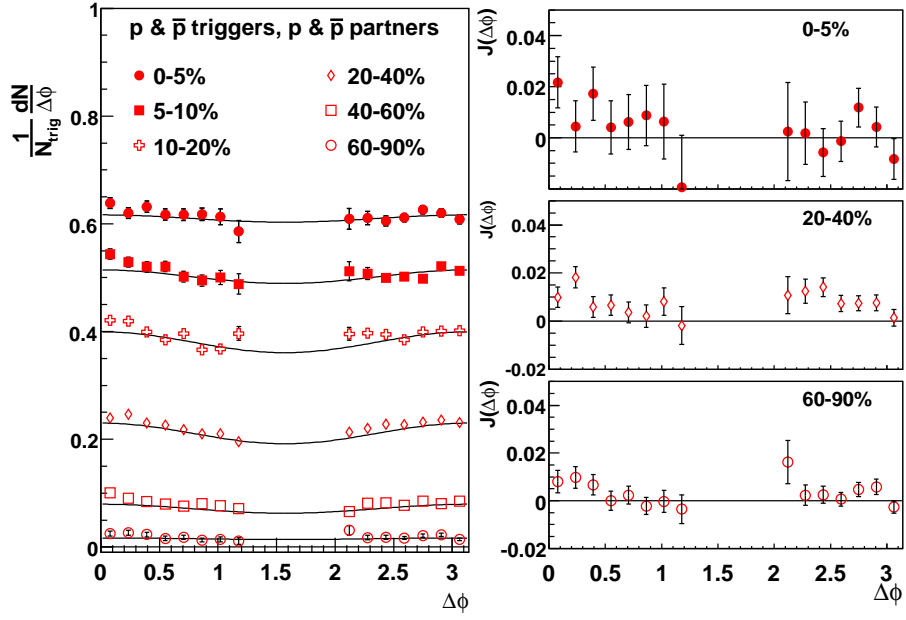}
\caption{Left: $\frac{1}{N_{trig}}\frac{dN}{d\Delta\phi}$ distributions
for charge-inclusive baryon triggers and associated particles for six centrality bins.
Triggers have 2.5~$<p_T<$~4.0~\gevc and
associated particles have 1.8~$<p_T<$~2.5~\gevc.
The solid lines indicate the combinatorial background modulated by 
elliptic flow.  Right:
Jet distributions, $J(\Delta\phi)$, after combinatorial background
and elliptic flow subtraction for 0-5\% (top), 20-40\% (middle) and 60-90\%
(bottom) centralities.
In all panels, only the statistical errors are shown. }
\label{dndphi_bar}
\end{figure*}

We first study baryon-baryon correlation, where both trigger and associated particles 
are identified as either $p$ or $\bar{p}$.
The left panel of Figure \ref{dndphi_bar} shows the azimuthal angular difference,
$\Delta\phi$, between charge inclusive $p$ and $\bar{p}$ measured in 
six centrality classes.   The solid lines show the combinatorial 
background level modulated by the expected correlation
due to elliptic flow, $B(1+2v_2^{trig}v_2^{assoc}\cos(2\Delta\phi))$. 
The excess is attributed to jet 
correlations, $J(\Delta\phi)$.  The azimuthal angular difference
distributions are then described by:
\begin{equation}
\frac{1}{N_{trig}}\frac{dN}{d\Delta\phi} = B(1 + 2v_2^{trig}v_2^{assoc}\cos(2\Delta\phi)) + 
J(\Delta\phi).
\label{eqncf}
\end{equation}
The region around $\Delta\phi=\pi/2$
has very limited  acceptance for pairs due to the requirement that
the trigger particle be measured in the TOF detector.  
The right panels of Figure \ref{dndphi_bar} show the $J(\Delta\phi)$
for three centralities after the combinatorial background subtraction.
There is a pronounced jet peak at small relative angles (near side), however,  there is no
visible structure on the away side, where the yields are slightly above or at the 
level of the combinatorial background.

\begin{figure*}[thb]
\includegraphics[width=0.7\linewidth]{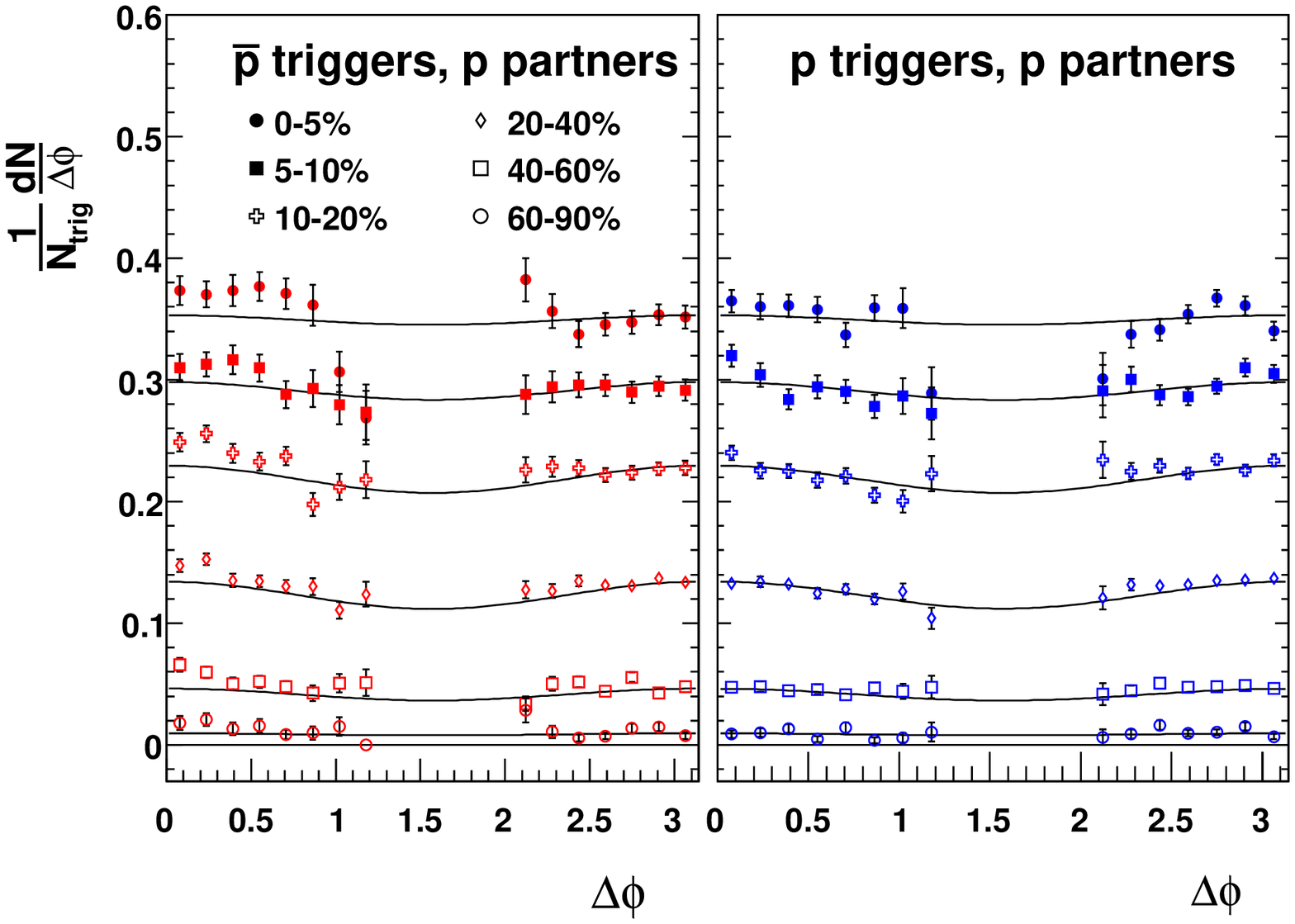}
\caption{$\frac{1}{N_{trig}}\frac{dN}{d\Delta\phi}$ distributions
for charge selected $\bar{p}$ (left) and $p$ (right) triggers 
both with associated $p$ for six centrality bins.  
Triggers have 2.5~$<p_T<$~4.0~\gevc and
associated particles have 1.8~$<p_T<$~2.5~\gevc.
The solid lines indicate the combinatorial background modulated by 
elliptic flow. Only the statistical errors are shown.}
\label{dndphi_p}
\end{figure*}

\begin{figure*}[thb]
\includegraphics[width=0.7\linewidth]{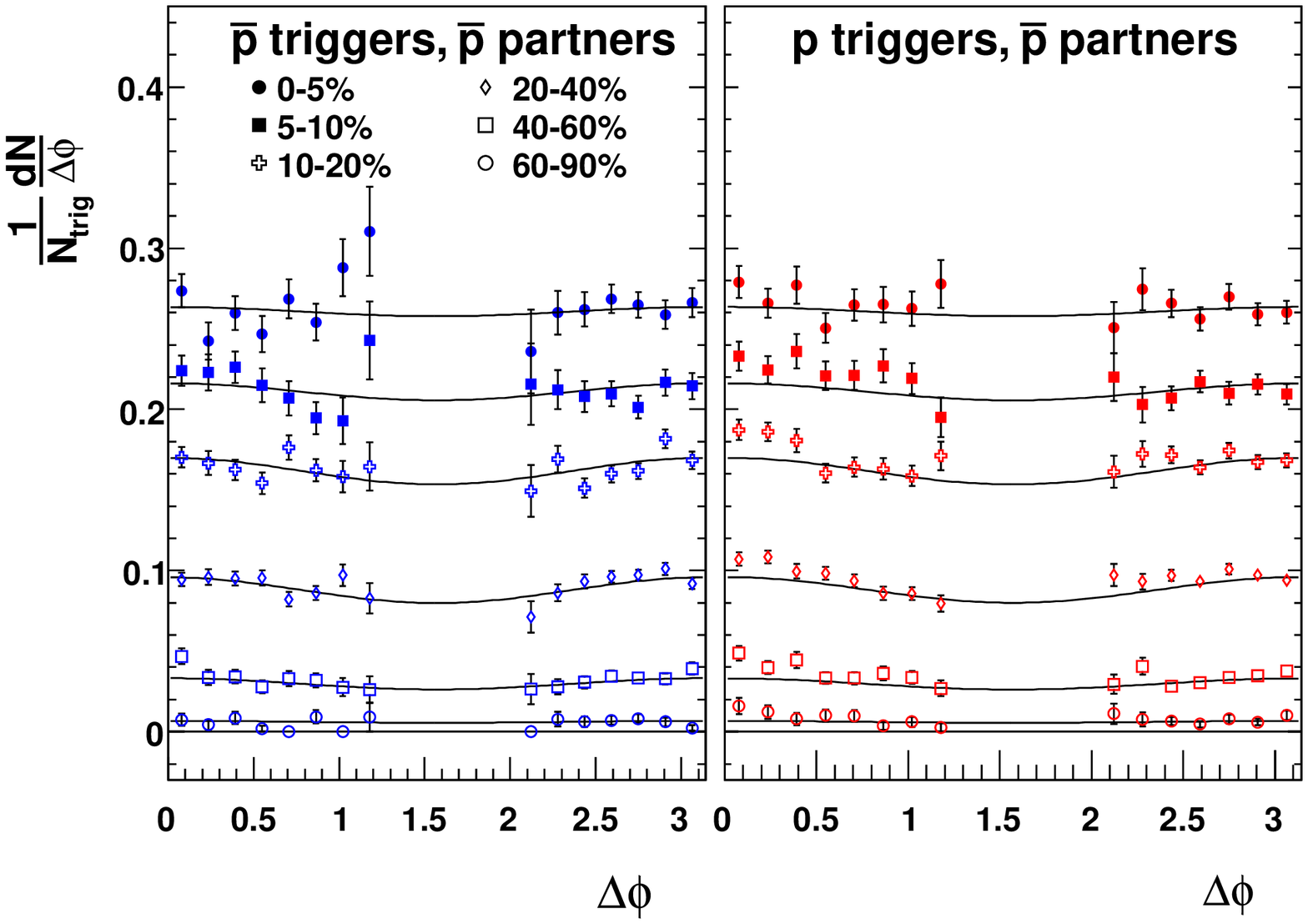}
\caption{$\frac{1}{N_{trig}}\frac{dN}{d\Delta\phi}$ distributions
for charge selected $\bar{p}$ (left) and $p$ (right) triggers 
both with associated $\bar{p}$
for six centrality bins.
Triggers have 2.5~$<p_T<$~4.0~\gevc and
associated particles have 1.8~$<p_T<$~2.5~\gevc.
The solid lines indicate the combinatorial background modulated by 
elliptic flow. Only the statistical errors are shown.}
\label{dndphi_pbar}
\end{figure*}

To further explore the observed structures, in 
Figures \ref{dndphi_p} and \ref{dndphi_pbar} we perform  the analysis
separately for each charge combination.  The correlations were measured 
in the same \pt range as in Figures \ref{dndphi_bar}.
A near side excess can be seen over the combinatorial background
for opposite sign pairs (left panel of Figure \ref{dndphi_p} and right panel
of Figure \ref{dndphi_pbar}) while no significant excess is seen for the same
sign pairs (right panel of Figure \ref{dndphi_p} and left panel of Figure \ref{dndphi_pbar}).

\begin{figure*}[thb]
\includegraphics[width=0.8\linewidth]{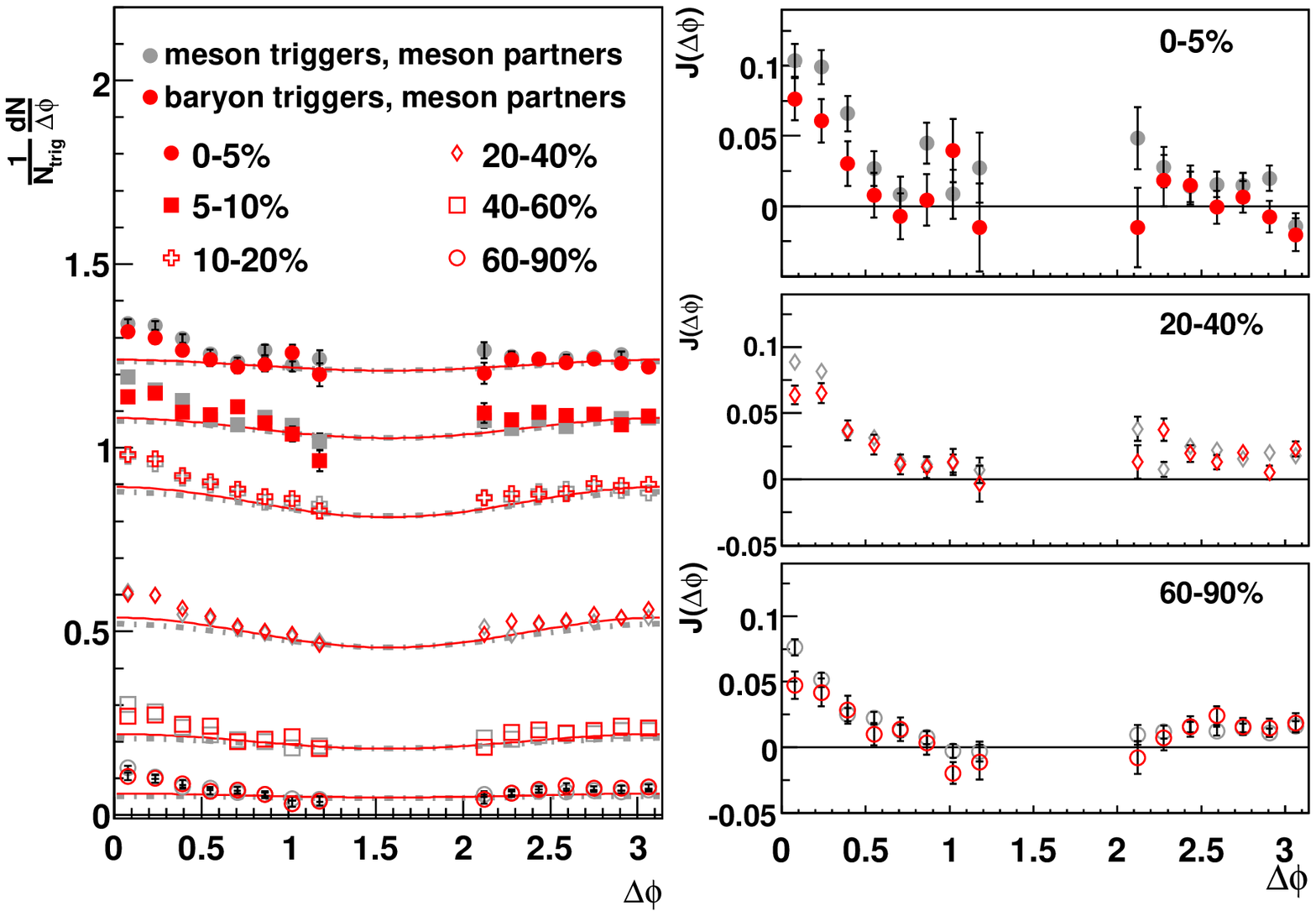}
\caption{Left:$\frac{1}{N_{trig}}\frac{dN}{d\Delta\phi}$ distributions
for charge-inclusive baryon and meson triggers and 
associated mesons for six centrality bins.
Triggers have 2.5~$<p_T<$~4.0~\gevc and
associated particles have 1.8~$<p_T<$~2.5~\gevc.
The solid lines indicate the combinatorial background modulated by 
elliptic flow. Right: Jet distributions, $J(\Delta\phi)$, after combinatorial
background and elliptic flow subtraction for 0-5\% (top), 20-40\% (middle) and
 60-90\% (bottom)
centralities.  In all panels, only the statistical errors are shown.}
\label{dndphi_mes}
\end{figure*}

The correlations involving mesons as associated particles provide a comparison baseline for the 
baryon-baryon correlations. We study both  baryon and meson
triggers associated with mesons.  
The left panel of Figure \ref{dndphi_mes} shows the baryon-meson and meson-meson correlations.
The right panels of Figure \ref{dndphi_mes}
show $J(\Delta\phi)$ for the 0-5\%, 20-40\% and 60-90\% centrality classes.
While in the mid-central 
collisions the meson and baryon triggered distributions agree well, 
the baryon triggered distributions 
in the most central collisions 
lie systematically below
the meson triggered points both before and after the combinatorial background
subtraction on the near side.

\begin{figure*}[thb]
\includegraphics[width=0.7\linewidth]{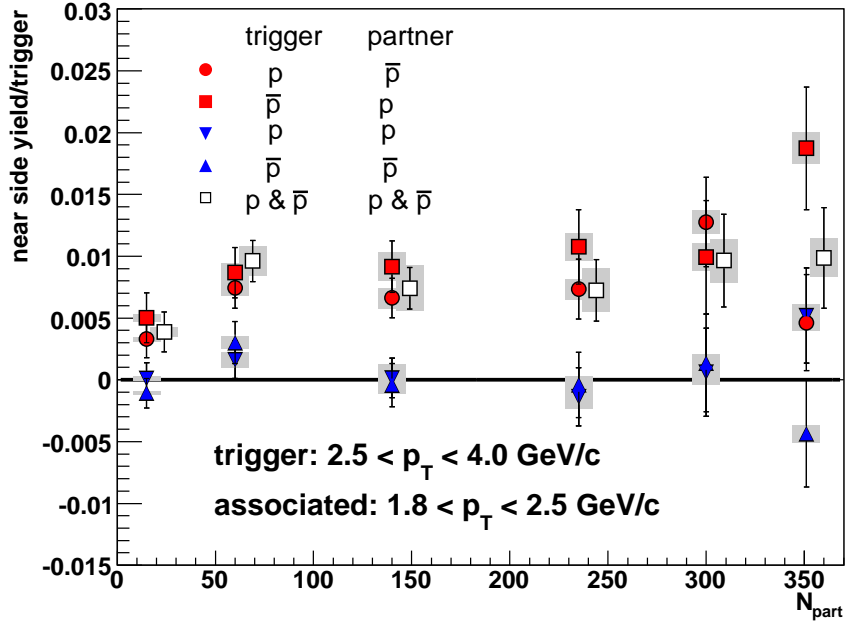}
\caption{Conditional yields per trigger  
on the near side for charge selected (solid points) and charge
selected (hollow points) $p$ and $\bar{p}$ correlations.
Triggers have
2.5~$<p_T<$~4.0~GeV/$c$ and associated particles have 1.8~$<p_T<$~2.5~GeV/$c$.
The error bars are the statistical errors and the boxes show the systematic
errors.
There is an 11.4\% additional normalization error on baryon associated particle points
and 8.9\% each on the $p$ and $\bar{p}$ associated particle points.}
\label{ppbar_near}
\end{figure*}

\begin{figure*}[hbt]
\includegraphics[width=0.7\linewidth]{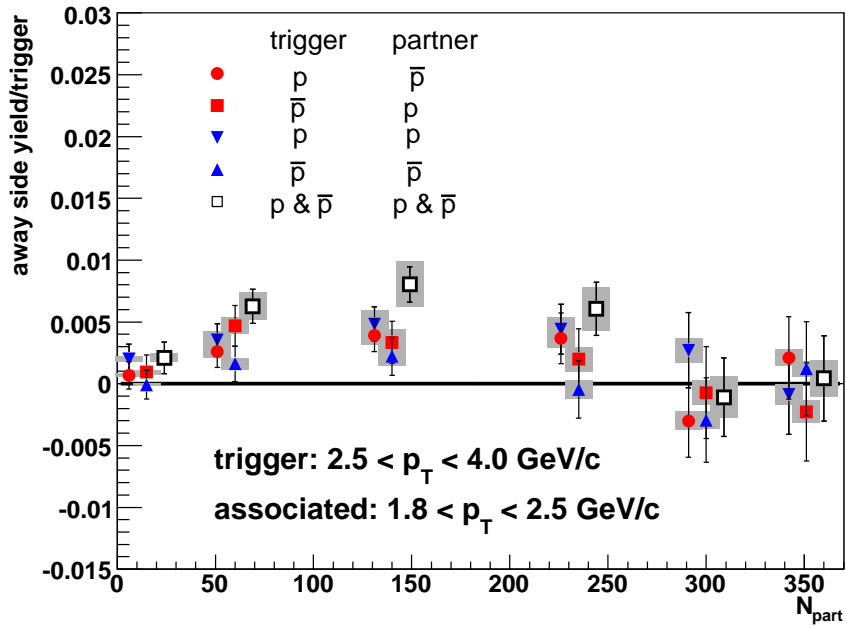}
\caption{Conditional yields per trigger  
on the away side for charge selected (solid points) and charge
selected (hollow points) $p$ and $\bar{p}$ correlations.
Triggers have
2.5~$<p_T<$~4.0~GeV/$c$ and associated particles have 1.8~$<p_T<$~2.5~GeV/$c$.
The error bars are the statistical errors and the boxes show the systematic
errors.
The additional normalization error is
the same as in Figure \ref{ppbar_near}.}
\label{ppbar_far}
\end{figure*}

To quantify the observed differences in the various trigger-associated particle combinations, we 
integrate the $J(\Delta\phi)$ distributions in the regions 0.0$<\Delta\phi<$0.94~rad and  $\pi-0.94<\Delta\phi<\pi$~rad to 
obtain the near-side and the away-side  conditional yields, respectively.  
Figures~\ref{ppbar_near} and~\ref{ppbar_far} show the conditional yield
per trigger as  a function of the number participating nucleons ($N_{part}$). The results were 
obtained from the data in Figures~\ref{dndphi_p} and \ref{dndphi_pbar} 
(solid points) and Figure~\ref{dndphi_bar} (open points) by integrating the $J(\Delta\phi)$  in 
the $\Delta\phi$ ranges specified above. This integration range excludes a large part of the 
away side shape modifications observed in \cite{ppg032}.  The small integration range
is used in this analysis only because of the limited acceptance
around $\Delta\phi=\pi/2$ due to the requirement to measure
the trigger particle in the TOF.
These results quantify the centrality
and particle type dependence of the jet-like correlations.
Figure~\ref{ppbar_near} shows that  the correlations between opposite sign
baryon pairs 

\clearpage 

\begin{figure*}[thb]
\includegraphics[width=0.8\linewidth]{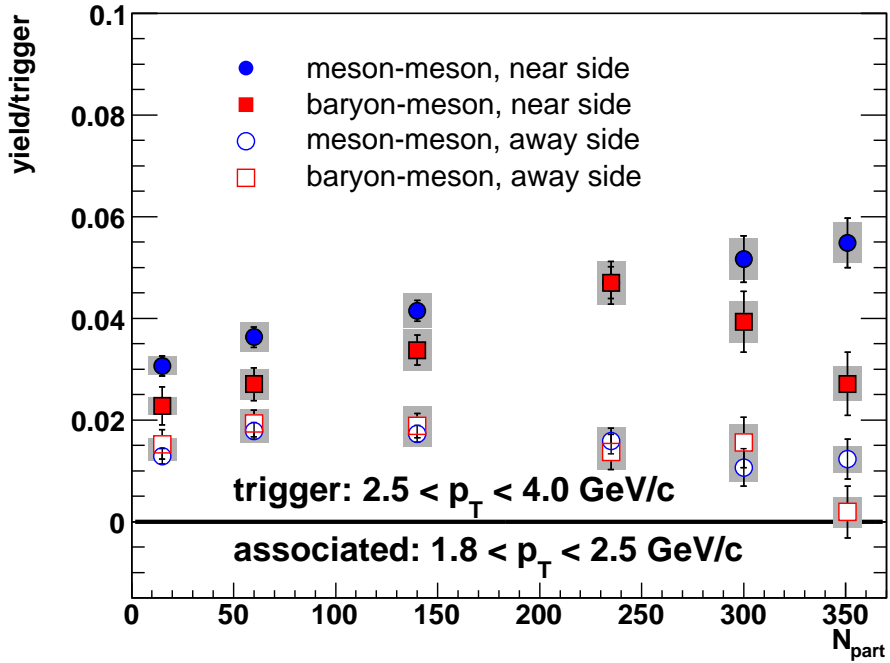}
\caption{Conditional yields per trigger for 
baryon (squares) and meson (circles) triggers with 
associated mesons. Triggers have
2.5~$<p_T<$~4.0~GeV/$c$ and associated particles have 1.8~$<p_T<$~2.5~GeV/$c$.
The error bars are the statistical errors and the boxes show the systematic errors.
There is an additional 13.6\% normalization error.}
\label{mes_near}
\end{figure*}

\noindent produce a significant non-zero conditional yield that is nearly independent
of centrality and that 
there is no significant yield associated with same sign pairs.
The open points in Figure \ref{ppbar_near}
 are from the charge inclusive analysis and show that
the conditional yield does come mainly from opposite sign pairs.
For all but the most peripheral point, no centrality dependence
is observed in the charge inclusive analysis.  The most peripheral
point sits 2.5$\sigma$ below the average conditional yield
for the other centrality bins.  The charge inclusive data are also consistent with
a linear increase with $N_{part}$.
  The systematic
errors on these distributions are highly correlated 
since the \flow values and centrality
correlation corrections are the same for all points at a given centrality.
There is an 8.6\% systematic error on the relative 
normalizations of the associated particle
$p$ and $\bar{p}$ points which is not shown.  
Figure~\ref{ppbar_far} shows the same correlations as Figure~\ref{ppbar_near}
for the away side region.  Here the charge inclusive points lie above the 
charge selected points for peripheral and mid-central collisions because
both same and opposite charge pairs have non-zero conditional yield.
The conditional yield rises from peripheral to mid-central collisions.  In
the most central collisions both the charge inclusive and charge selected
points are consistent with zero.

Figure~\ref{mes_near} shows both the near and away side conditional
yield for baryon triggers and associated mesons
 and meson triggers with associated mesons.
The meson trigger-associated mesons 
 conditional yield on the near side rises smoothly with centrality.
The baryon triggered yields are systematically
lower than the meson triggered yields, but
 also rise linearly with $N_{part}$ for $N_{part}<$~250.
In central collisions
 the baryon triggered yields are lower 
than the linear $N_{part}$ dependence observed
in the meson-meson points for all $N_{part}$ 
and the baryon-meson points for $N_{part}<$~250.
The statistical errors on baryon-meson conditional yields exclude a linear
increase with $N_{part}$ on the 2.6$\sigma$ level.  The
systematic errors on the \flow values and the centrality correlation
correction are correlated with centrality. 
The most central baryon triggered point is consistent with
the most peripheral baryon triggered point.  The most central
meson-meson conditional yield is 70\%~$\pm$~20\% greater than in
peripheral collisions.  On the away side, no significant dependence on trigger
type is observed.

\section{Discussion}
\label{section:discussion}
The observed $x_T$ scaling and pQCD calculations suggest hard scattering is
the dominant mechanism for particle production for $p_T >$2~\gevc
at $\sqrt{s}$=200~GeV in p+p collisions~\cite{starxt}.   $\pi^0$ production in Au+Au collisions
at $\sqrt{s_{NN}}$~=~200~GeV follows $x_T$ scaling \cite{ppg023} and is well
described by perturbative theories which include radiative energy
loss by hard partons traversing the medium~\cite{vitev,wang,eskola}
at \pt as low as 2~\gevc.
Models which describe the excess of baryons relative to mesons \cite{ppg015}
typically do so by some mechanism which extends the \pt range of soft physics.
Here we have used two particle azimuthal correlations to
study the particle type dependence of jet-like correlations
in Au+Au collisions in the region of the baryon excess.
Since jets in $e^+ + e^-$ collisions 
fragment dominantly into mesons~\cite{delphi}, we
take meson-meson correlations as a baseline for jet
fragmentation in Au+Au collisions.  
The increase in meson-meson
near side conditional yield with centrality seen in Figure \ref{mes_near}
has also been observed in meson-hadron correlations \cite{ppg033}
and hadron-hadron correlations \cite{starb2b,ppg032} and
is not yet quantitatively understood.  Here we are interested 
primarily in jet correlations of baryons so meson-meson 
correlations provide a useful reference.
The yield of associated mesons
per trigger baryon is systematically lower than the yield of
associated mesons per trigger meson, but  
baryon-meson correlations on the near side increase as meson-meson 
correlations for all centralities
except for the most central.  
On the away side there is no significant dependence on
the trigger type for associated mesons, 
as is expected if baryon and meson triggers come from
jets of  approximately the same energy
 and if the di-jets fragment independently of the
trigger jet.
The yield of associated baryons  per baryon trigger on the
near side is observed to be nearly constant with centrality, except
for the most peripheral point which is significantly lower than the others.  
The data are also consistent with a linear increase in the conditional
yield as a function of$N_{part}$.  The small yield of associated
baryons per baryon trigger does not imply that baryon number is not
conserved within the near side jet
since the \pt range of the measured associated particles is narrow,
 and the PHENIX $\eta$ acceptance does not contain all of the
associated particles. 

The data presented here are consistent with baryons at 2.5~$<p_T<$~4.0~\gevc  
arising predominately from hard scattering processes.  
First, the yield of mesons  associated with baryon triggers
has the same centrality dependence as associated  mesons 
per meson trigger for $N_{part}<$~250 despite a change in
the $\bar{p}/\pi^-$ ratio by a factor of three from peripheral
collisions.  Second, the away side yield into 0.94~rad is independent of
the trigger type, consistent with the away side jet fragmenting
independently of the trigger jet.  Lastly, the charge dependence 
of $p$ and $\bar{p}$ correlations show that small angle $p$-$\bar{p}$
pairs are correlated beyond the expected correlations from
elliptic flow, and that small angle $p$-$p$ and $\bar{p}$-$\bar{p}$
pairs are not.  This is true in peripheral collisions where the $\bar{p}/\pi^-$
ratio is close to the value from p+p collisions \cite{ppg030}
and also in central collisions where the ratio is a factor of three
larger. This indicates that the mechanism producing the baryon excess
is also producing small angle $p$-$\bar{p}$ pairs.

The results for near side conditional yields presented here 
disagree with the recombination model calculation in
\cite{friesmit} which predicts a very weak centrality 
dependence for meson-meson and baryon-meson conditional yields
and nearly the same magnitude for  baryon-meson and baryon-baryon
near side conditional yields. In contrast, the data show 
the conditional yield of associated mesons 
with baryon triggers to be a factor of two
to five times larger than the conditional yield of  baryons
associated with baryon triggers, depending on centrality.
 The results presented here also appear to exclude
baryon production via higher twist mechanisms \cite{brodsky} which would produce
isolated $p$ and $\bar{p}$.
No correlation calculations are available from
the gluon junction model~\cite{topor-pop}, so a comparison 
beyond the successfully described single particle data could 
not be done at this point.

We have systematically explored the particle type dependence of
jet fragmentation at intermediate \pt in Au+Au collisions
at $\sqrt{s_{NN}}$~=~200~GeV.  
The new data disagree with calculations from the 
recombination model presented in \cite{friescs,friesmit}.
Given the success of recombination models in 
reproducing elliptic flow and hadron spectra
data it would be interesting to see if  other recombination
calculations are able to describe the data presented here.
We find that near side correlations between meson triggers and 
associated mesons increase with centrality.  Near side
correlations between baryon triggers and associated mesons show the 
same centrality dependence except for the most central collisions
where there is a significant decrease.  The first measurements of
baryon pairs on the near side are found to be largely due to opposite 
charge $p$-$\bar{p}$ pairs.  
Under the assumption that the above centrality dependencies
of particle pairs and single particles are not coincidental,
one can explain the observed baryon excess at intermediate \pt
in Au+Au collisions via jet induced production of
baryon-antibaryon pairs.



We thank the staff of the Collider-Accelerator and 
Physics Departments at BNL for their vital contributions.  
We acknowledge support from 
the Department of Energy and NSF (U.S.A.), 
MEXT and JSPS (Japan), 
CNPq and FAPESP (Brazil), 
NSFC (China), 
MSMT (Czech Republic),
IN2P3/CNRS, and CEA (France), 
BMBF, DAAD, and AvH (Germany), 
OTKA (Hungary), 
DAE (India), 
ISF (Israel), 
KRF and KOSEF (Korea), 
MES, RAS, and FAAE (Russia),
VR and KAW (Sweden), 
U.S. CRDF for the FSU, 
US-Hungarian NSF-OTKA-MTA, 
and US-Israel BSF.

\label{}




\bibliographystyle{elsart-num}






\end{document}